\documentclass[a4paper]{jpconf}
\usepackage{graphicx}

\begin{document}

\title[Exciton formation and dissociation]{Exciton formation and dissociation in
mass-asymmetric electron-hole plasmas}

\author{H.~Fehske$^1$, V.~Filinov$^{2,3}$, M.~Bonitz$^3$, V.~Fortov$^2$ and P.~Levashov$^2$
}
\address{$^1$Institut f\"ur Physik, Ernst-Moritz-Arndt-Universit{\"a}t
Greifswald, Domstrasse 10a, D-17489 Greifswald, Germany}
\address{$^2$
Institute for High Energy Density, Russian Academy of Sciences,
Izhorskay 13/19, Moscow 127412, Russia
}
\address{$^3$Christian-Albrechts-Universit\"at zu Kiel, Institut f\"ur
Theoretische Physik und Astrophysik, Lehrstuhl Statistische
Physik, Leibnizstrasse 15, 24098 Kiel, Germany }

\begin{abstract}
First-principle path integral Monte Carlo simulations were
performed in order to analyze correlation effects in complex
electron-hole plasmas, particularly with regard to the appearance
of excitonic bound states. Results are discussed in relation to
exciton formation in unconventional  semiconductors with large
electron hole mass asymmetry.
\end{abstract}




\section{Introduction}
Strongly coupled Coulomb systems have been in the focus of recent
investigations in many fields, including plasma, nuclear,
condensed matter and astrophysics~\cite{boston97,B1a,B1b,B1c}. In
these systems the Coulomb energy is of the order of, or even
exceeds the mean kinetic energy. Besides, quantum effects are of
vital importance. The formation of Coulomb liquids and solids is
only one of the multi-faceted phenomena that might show up. Just
now the existence of Coulomb crystals has been studied for neutral
plasmas containing (at least) two oppositely charged
components~\cite{bfflf05}. Most notably crystallization of holes
in semiconductors was predicted to occur in materials with a
sufficiently large hole to electron effective mass asymmetry of
about 80.

Since bound state formation turned out to be the relevant limiting
factor for the appearance of such an exceptional quantum hole
(Coulomb) crystal at low temperatures~\cite{bfflf05}, the aim of
the present contribution is to analyze the formation of excitons
in mass asymmetric electron-hole plasmas from first principles. A
theoretical approach that is well suited to address this problem
is the direct path integral quantum Monte Carlo (DPIMC) method for
the N-particle density operator~\cite{FiBoEbFo01}. The DPIMC
technique avoids additional approximations, such as the fixed node
and restricted path integral method. It has been successfully
applied to treat both strong interaction and quantum
effects~\cite{FiBoEbFo01,Fetal,afilinov-etal.00pss}.

The organization of the paper is as follows. In Sec. 2 we briefly
describe the essence of the DPIMC method we employed. Section 3
presents our simulation results. There we will give a detailed
discussion of typical many-particle configurations at low and high
temperatures, as well as of electron-electron, hole-hole and
electron-hole pair distribution functions and structure factors.
We conclude in Sec.~3, relating our results to the heavily debated
exciton formation in intermediate valent Tm[Se,Te]~\cite{wachter}.
\section{Direct path integral Monte Carlo approach}
The starting point for the description of a two-component
(equilibrium) quantum plasma is the partition function given at
inverse temperature $\beta=1/k_B T$ by
\begin{equation}\label{q-def}
Z(N_e,N_h,V,\beta) = \frac{1}{N_e!N_h!} \sum_{\sigma}\int\limits_V
dq \,\rho(q, \sigma ;\beta),
\end{equation}
where
$q=\{q_e,q_h\}$ and $\sigma=\{\sigma_e,\sigma_h\}$
denote the spatial coordinates and spin degrees of freedom
of $N_e$ electrons and $N_h$ holes ($N_e=N_h$), respectively,
i.e. $q_a=\{q_{1,a}\ldots q_{l,a}\ldots q_{N_a,a}\}$ and
$\sigma_a=\{\sigma_{1,a}\ldots \sigma_{l,a}\ldots  \sigma_{N_a,a}\}$
with $a=e,h$. Then the pair distribution functions
and the charge structure factors
can be expressed as~\cite{ifisher}
\begin{equation} \label{gab-rho}
g_{ab}(r) = \frac{N_e!N_h!}{Z(N_e,N_h,\beta)}
\sum_{\sigma} \int\limits_{V} dq
\,\delta(r_{1,a}-q_{1,a})\, \delta(r_{2,b}-q_{2,b})
\,\rho(q,\sigma;\beta)
\end{equation}
and
\begin{eqnarray}
S_{ab}(k) &=& \frac{\int_{0}^{\infty}dr
r^2[g_{ab}(r)-1]\sin(kr)/(kr)}{|\int_{0}^{\infty}dr
r^2(g_{ab}(r)-1)|}\,, \label{sab}
\end{eqnarray}
where $r=r_{1,a}-r_{2,b}$.
\begin{figure}[b]
\begin{minipage}{0.65\linewidth}
\hspace{1cm}
\includegraphics[width=6cm,clip=true]{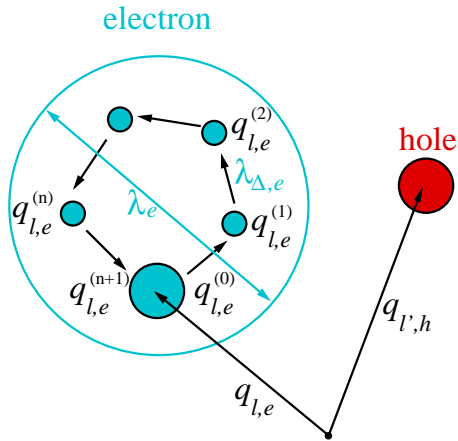}
\end{minipage}
\begin{minipage}{0.35\linewidth}
\vspace*{3cm} \caption{(Color online) Bead representation of
electrons and holes. Here $\lambda_e^2=2\pi\hbar^2\beta/m_e$,
$\lambda_{\Delta,e}^2=2\pi\hbar^2\Delta\beta/m_e$,
$q^{(1)}_{l,e}=q^{(0)}_{l,e}+\lambda_{\Delta,e}\,\xi^{(1)}_{l,e}$,
and $\sigma=\sigma^{\prime}$. The holes have a similar beads
representation, however $\lambda _h$ is $\sqrt{m_e/m_h}$ times
smaller.} \label{beads}
\end{minipage}
\end{figure}

Of course, in general the exact density matrix
$\rho(q, \sigma;\beta)$ of an interacting quantum many-particle
systems is not known but can be constructed using a path integral approach:
\begin{eqnarray}
&&\hspace*{-2cm}\int\limits_{V}\sum_{\sigma} dq^{(0)}\,
\rho(q^{(0)},\sigma;\beta)=
\int\limits_{V} dq^{(0)} \dots dq^{(n)} \,
\rho^{(1)}\cdot\rho^{(2)} \, \dots \rho^{(n)}
\nonumber\\
\hspace*{1cm}&\times&\sum_{\sigma}\sum_{P_e} \sum_{P_h} (\pm 1)^{\kappa_{P_e}+
\kappa_{P_h}} \,{\cal S}(\sigma, {\hat P_e} {\hat P_h} \sigma_{a}^\prime)\,
{\hat P_e} {\hat P_h} \rho^{(n+1)}_{ q_a^{(n+1)}=
q_a^{(0)}, \sigma_a'=\sigma_a}\,.
 \label{rho-pimc}
\end{eqnarray}
Here $\rho^{(i)}\equiv
\rho\left(q^{(i-1)},q^{(i)};\Delta\beta\right) \equiv \langle
q^{(i-1)}|e^{-\Delta \beta {\hat H}}|q^{(i)}\rangle$ and $\Delta
\beta \equiv \beta/(n+1)$. The Hamiltonian  for the electron-hole
plasma (${\hat H}={\hat K}+{\hat U}$) contains kinetic energy
(${\hat K}$) and Coulomb interaction energy (${\hat  U} = {\hat
U}_{ee} + {\hat  U}_{hh} + {\hat  U}_{eh}$) parts. In
Eq.~(\ref{rho-pimc}) the index $i=1\dots n+1$ labels the
high-temperature $[(n+1)k_B T]$ density matrices $\rho^{(i)}$.
Accordingly each particle is represented by $(n+1)$ beads, i.e.
the whole configuration of the particles is represented by a
$3(N_e+N_h)(n+1)$--dimensional vector
$\tilde{q}\equiv\{q_{1,e}^{(0)}, \dots q_{1,e}^{(n+1)},
q_{2,e}^{(0)}\ldots q_{2,e}^{(n+1)}, \ldots q_{N_e,e}^{(n+1)};
q_{1,h}^{(0)}\ldots q_{N_h,h}^{(n+1)} \}$. Further details of the
high-temperature path integral representation are given in
Ref~\cite{FiBoEbFo01}.

Figure.~\ref{beads} illustrates the representation of one (light)
electron and one (heavy) hole. The circle around the electron
beads symbolizes the region that mainly contributes to the
partition function path integral. The size of this region is of
the order of the thermal electron wavelength $\lambda_e(T)$, while
typical distances  between electron beads are of the order of the
electron wavelength taken at $(n+1)$-times higher temperatures.
The same representation is valid for holes but due to the larger
hole mass the characteristic length scales are smaller by an
factor of $\sqrt{m_e/m_h}$. The spin gives rise to the spin part
of the density matrix (${\cal S}$) with exchange effects accounted
for by the permutation operators  $\hat P_e$ and $\hat P_h$ acting
on the electron and hole coordinates $q^{(n+1)}$ and spin
projections $\sigma'$. The sum is over all permutations with
parity $\kappa_{P_e}$ and $\kappa_{P_h}$.

In the evaluation of the expressions obtained so far, we make use
of a Monte Carlo scheme~\cite{zamalin} with different types of
Monte Carlo steps: Either electron or hole coordinates ($q_{l,a}$)
or dimensionless individual electron or hole beads
($\xi_{l,a}^{(i)}$) were moved until convergence is reached.
Periodic boundary conditions are applied to the basic Monte Carlo
cell. The procedure has been extensively tested. For example, the
comparison with the known analytical expressions for pressure and
energy of an ideal Fermi gas showed that Fermi statistics is well
reproduced~\cite{FiBoEbFo01}. Moreover we applied the method to
few-electron systems in harmonic traps where again the
analytically known limiting behavior of the energy was
recovered~\cite{afilinov-etal.00pss}. For the present simulations
of an mass-asymmetric electron-hole plasma, we have varied both
particle and beads numbers and found that in order to obtain
convergent results particle numbers $N_e=N_h= 50$ and beads
numbers in the range of $N_b=20$ are sufficient. Thereby the
density matrix in the high temperature decomposition always was
taken at temperatures above the electron-hole binding energy.
Finally let us emphasize that the maximum statistical error of the
results presented below is about $5\%$ and can be systematically
reduced by increasing the length of the Monte Carlo run.
\section{Simulation results}
We now use the DPIMC scheme outlined in the preceding section in
order to study exciton  formation in (non-ideal) electron-hole
plasmas. Independent model parameters are the temperature ($T$),
the masses of electrons ($m_e$) and holes  ($m_h$), and the static
dielectric constant ($\varepsilon$). The plasma density is
characterized by the Brueckner parameter $r_s$, defined as the
ratio of the mean distance between the particles $d=[3/(4\pi
(n_e+n_h))]^{1/3}$ and the Bohr radius $a_B$ \cite{note1}, where
$n_e$ and $n_h$ are electron and hole densities, respectively. To
make contact with experiments on the Tm[Te,Se]
system~\cite{wachter}, we choose  $m_e=2$, $m_h=80$, and
$\varepsilon =25$ (the masses are in units of the free electron
mass). Then, assuming a simple Wannier exciton picture, the
exciton binding energy would be about 500~K.

\subsection{Particle configurations}
\begin{figure}[t]
\begin{center}
\includegraphics[width=6cm,clip=true]{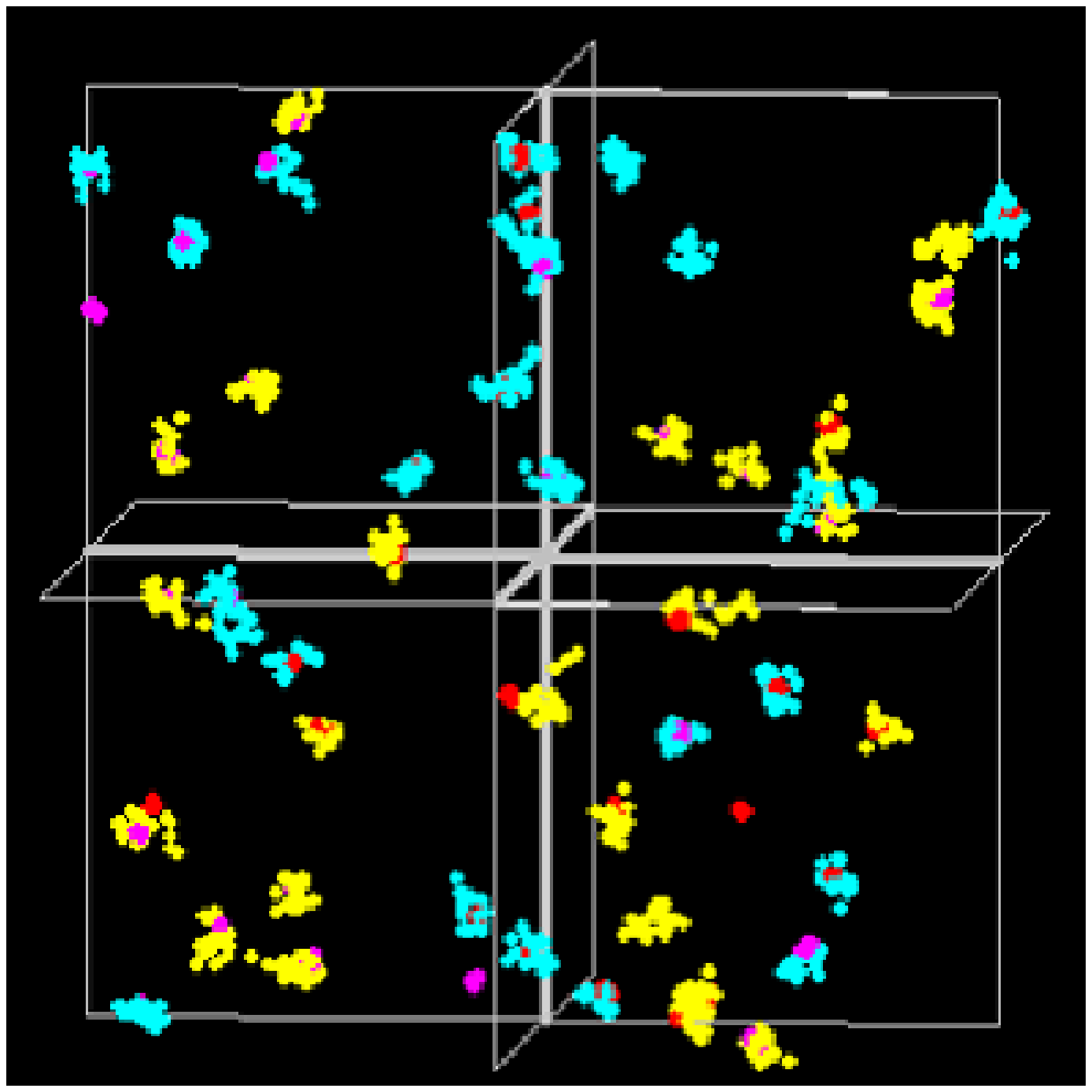} \hspace*{1cm}
\includegraphics[width=6cm,clip=true]{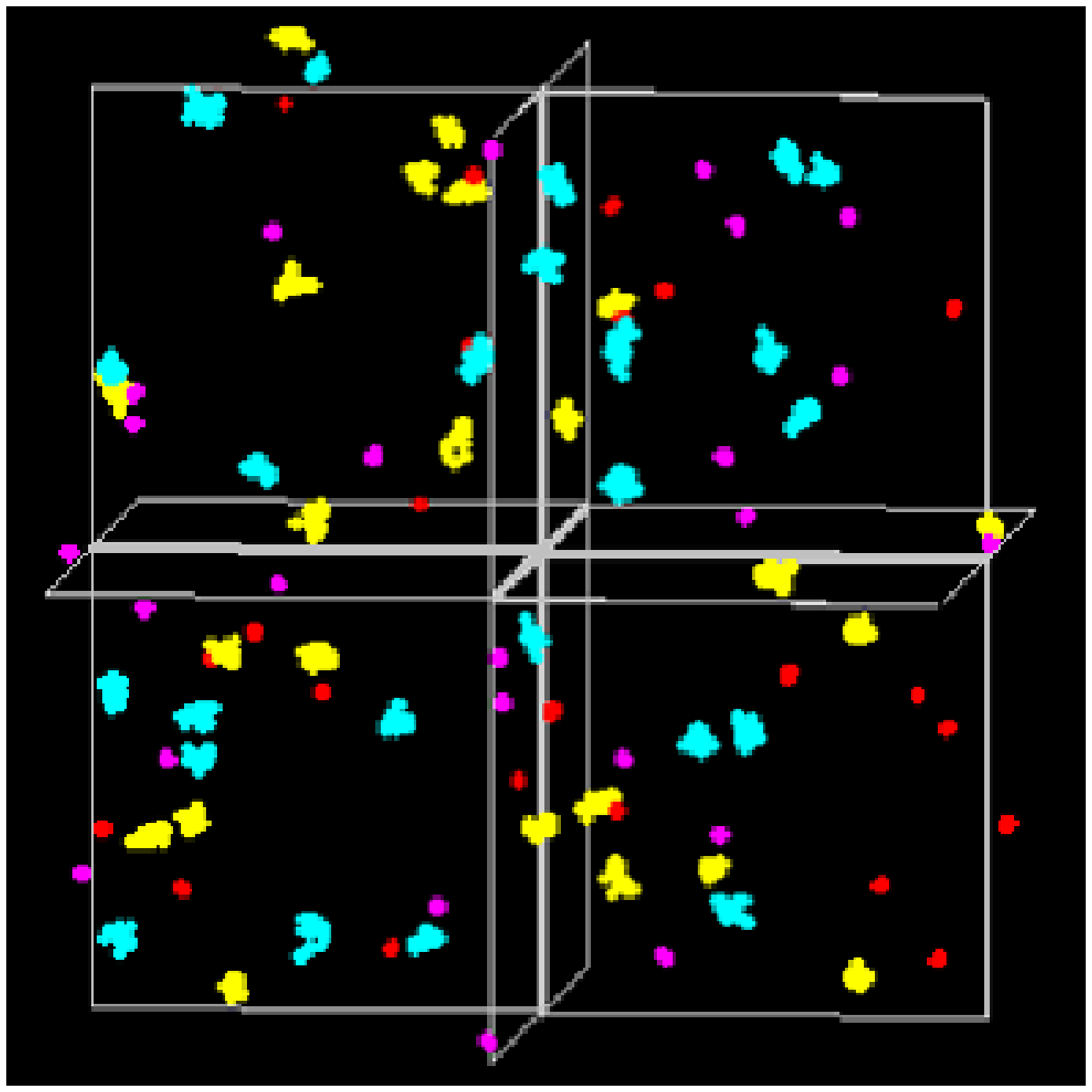}
\caption{(Color online) Snapshots of the electron-hole plasma at
$T = 50 K$ (left) and $T = 200 K$ (right) (for $r_s=10$). Clouds
of blue (yellow) dots represent electrons with spin up (down),
whereas red (pink) dots mark the holes. Grey lines indicate the
main simulation box which is periodically repeated in all spatial
directions.} \label{Snpsht2}
\end{center}
\end{figure}

Figure~\ref{Snpsht2} shows typical ``snapshots'' of an
electron-hole many-particle state with $r_s=10$ at low $(50 K)$
and high $(200 K)$ temperatures. Here, as a result of the
temperature decomposition of the density matrix, each electron and
hole is represented by several beads. Since the number of beads is
twenty the high temperature density matrices correspond to a
temperature of at least $1000 K$, being about two times larger
than the characteristic (excitonic) binding energy ($E_X^b$).

The spatial distribution of the beads reflects the position
probability amplitude of the many-particle wave function, where
the characteristic size of a certain cloud of beads is of the
order of the thermal wave length of a single quantum particle.
Because of the mass asymmetry the typical size of the electron
clouds is about $6$ times larger than that of the holes. Also,
increase of the temperature leads to a reduce of the particle
extension.  Interestingly, at low temperatures practically all
holes are closely covered by electron beads. This means electrons
and holes form predominantly bound states and the whole system
consists mainly of excitons. From Fig.~\ref{Snpsht2} we can see
that the average distance between particles (unbound electrons,
holes and excitons) is much larger than the size of the excitons.
At higher temperatures ($T=200 K $) we observe a significant
number of free electrons and holes (cf. Fig.~\ref{Snpsht2}, right
panel). Due to the temperature-induced dissociation of the
excitonic bound states a partially ionized (non-ideal)
electron-hole plasma is formed.
\begin{figure}[h]
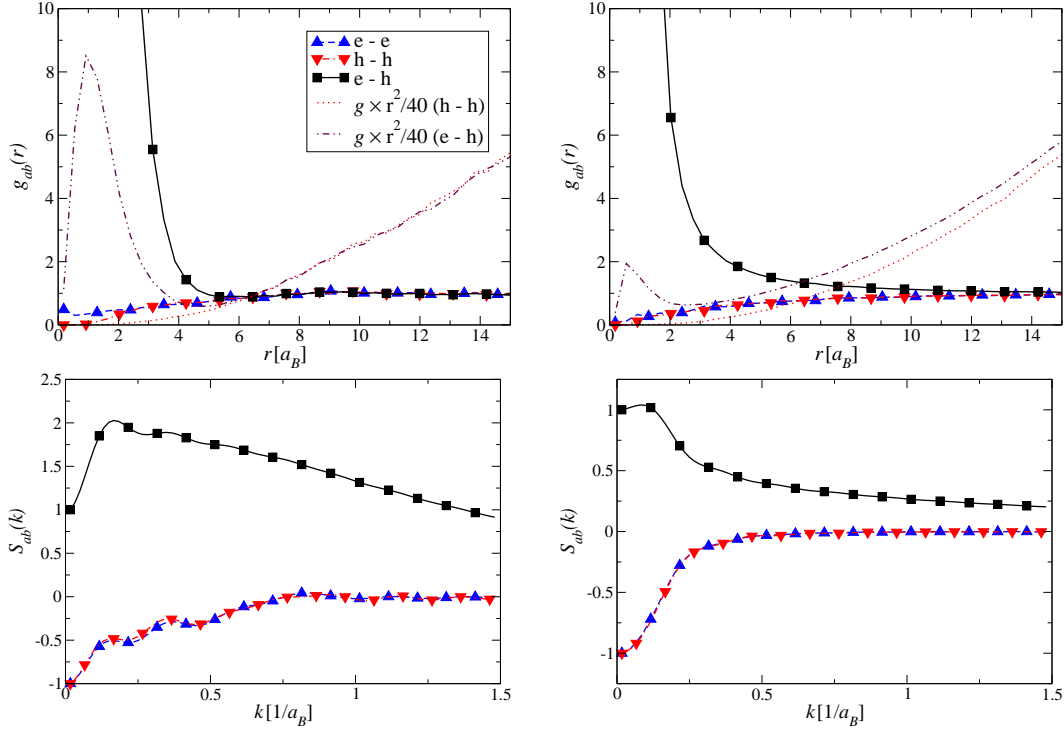

\begin{center}
\includegraphics[width=6.7cm,clip=true]{fig3a.eps}\hspace*{.5cm}
\includegraphics[width=6.7cm,clip=true]{fig3b.eps}\\
\includegraphics[width=6.7cm,clip=true]{fig3c.eps}\hspace*{.5cm}
\includegraphics[width=6.7cm,clip=true]{fig3d.eps}
\caption{(Color online) Pair distribution functions (upper panels)
and structure factors  (lower panels) at $T=50K$ (left panels) and
$T=200 K$ (right panels), respectively. Results are given for
$r_s=10$.}
 \label{geehh}
\end{center}
\end{figure}
\subsection{Pair distribution functions and structure factors}
This scenario is supported by the behavior of the
electron-electron, hole-hole and electron-hole correlations
pointed out in this section. The pair distribution functions and
structure factors shown in Fig.~\ref{geehh} are the sum over both
spin projections of the particles.

Let us first discuss the physics at short distances in terms of
the pair distribution functions. The function $g_{ab}$ is the
distribution of b-particles, on the average, about any a-particle.
As an effect of the Coulomb and Fermi (statistics) repulsion,
$g_{ee}$ and $g_{hh}$ are strongly depleted at small $r$. The
decay of the hole-hole correlations is much stronger because of
the larger masses, i.e. holes behave more ``classically''. For the
electron subsystem quantum exchange and tunnelling are more
important and compete with the Coulomb repulsion. Furthermore the
pronounced peak of the electron-hole pair distribution
unambiguously signals the formation exciton bound states. This is
confirmed by considering the product $r^2 g_{eh}(r)$ which has the
physical meaning of the probability to find the electron at
distance $r$ away from the hole. At low temperatures, $r^2
g_{eh}(r)$ is
strongly peaked around $a_B$. 
Increasing the temperature the excitonic peak weakens and finally
it vanishes at about 300~K, which reflects thermal dissociation of
excitons. Note that $g_{ee}$ and $g_{hh}$ (almost) coincide even
at short distances for 200~K, i.e. the now electrons behave more
like ``classical'' particles as well. Clearly all functions
$g_{ab}$ go to unity at large distances for all $T$, which is a
result of the overall uniform distributions of the charges.

To discuss the physical behavior at large distances in more
detail, it is convenient to consider the (charge) structure factor
$S_{ab}$ in momentum space. According to Eq.~(\ref{sab}) positive
(negative) values of $S_{ab}$ indicate attraction (repulsion).
From $g_{eh}$ the typical length scale of the attractive
electron-hole correlations was found to be of the order of $5a_B$
at 50~K (cf., Fig.~\ref{geehh}, upper left panel). Exciton
formation sets a new length scale for the structure factor as well
(cf. the maximum in the structure factor at about~0.2~$[1/a_B]$).
In accordance with the above discussion this excitonic maximum is
shifted to smaller $k$ values and finally washed out if the
temperature is raised. In the case that the maximum (minimum) of
$S_{ab}$ is located at $k=0$ and $S_{ab}(k)$ is a monotonically
decreasing (increasing) and rather structure-less function of~$k$;
the system closely resembles a homogeneous electron-hole plasma.
Nontrivial quantum (screening) effects, however, come into play at
low temperatures as can be seen from the wiggly behavior
especially of the electron-electron and hole-hole charge
correlations.

\subsection{Exciton density}
From the analysis of the snapshots in Sec.~3.1 we found that our
complex electron-hole plasma contains at the same time free
electrons, holes and excitons. That means, depending  on the
temperatures and particle densities, electrons and
holes may appear in bound and unbound quantum and classical
states. In a strict sense, we cannot calculate the fraction of
bound and free states in our DPIMC simulations because of the
possible overlapping of the electron ``shells'', in particular at
intermediate temperatures. Nevertheless a rough estimation of the
fraction of the number of electron-hole bound states can be
obtained by the following physical reasonings.

The contribution of all bound and scattering states at temperature
$1/\beta$ is given by  the two-particle Slater sum
\begin{equation}
\Sigma_{eh}(r,\beta)=8\pi
^{3/2}\lambda_{eh}^{3}\sum_{E_{\alpha}=E_{0}}^{\infty}|\Psi_{\alpha}(r)|^{2}\exp(-\beta
E_{\alpha})\,, \label{slsm}
\end{equation}
where $E_{\alpha}$ and $\Psi_{\alpha}(r)$ are the energy (without
the center of mass energy) and the wave function of the exciton,
respectively. $\Sigma_{eh}$ is, in essence, the diagonal part of
the density matrix and the product $r^2\Sigma_{eh} $ represents
the (quantum mechanical) probability for an electron and a hole to
be separated by distance $r$.

The sum over all possible states contains contributions from the
discrete (d) and continuum (c) part of the spectrum,
$\Sigma_{eh}=\Sigma^{d}+\Sigma^{c}$. $\Sigma^{d}$ contains all
populated bound states between $E_0$ and energy $E^\prime$ that
separates bound and free (scattering) states. For low densities it
is reasonable to choose $E' \approx - \frac{3}{2\beta}$ since
higher lying bound states will be thermally
ionised~\cite{zamalin}.

At temperatures smaller than the electron-hole binding energy and
distances smaller or of the order of several Bohr radii, the main
contribution to the Slater  sum comes from electron-hole bound
states, while at larger distances free electron-hole scattering
states turn out to be the crucial factor. We found that the
product $r^2\Sigma_{eh}$ is sharply peaked around $a_B$, just like
$r^2 g_{eh}(r)$ in Fig.~\ref{geehh}, where the maximum is most
pronounced at low temperatures.
\begin{figure}[b]
\begin{center}
\includegraphics[width=8cm,clip=true]{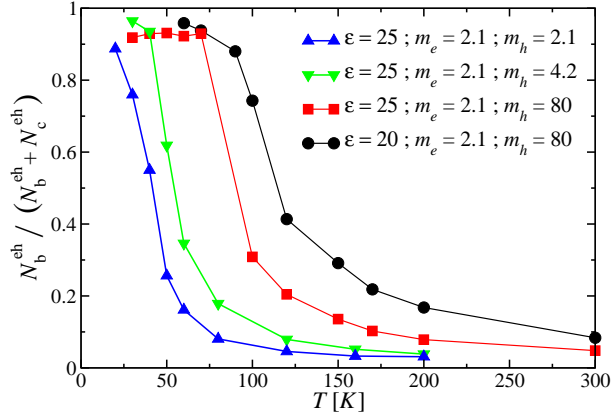}
\caption{(Color online) Fraction of electron-hole bound states
versus temperature for different hole masses and dielectric
constants.} \label{cmpexp}
\end{center}
\end{figure}
Increasing the temperature, $r^2 g_{eh}(r)$ approaches the
parabola $r^2$ related to the case of full absence of bound states
(ideal plasma $g_{eh}(r)\equiv 1$). An attractive interaction
results in an increase of the pair distribution function at short
e-h distances: $g_{eh}(r) > 1$ for $0\leq r\leq r^b$. Since the
total number of  particles is conserved this increase involves a
decrease at large distances (see Fig.~\ref{geehh}). So the
interval $[0,r^b]$ with $r^2*( g_{eh}(r)-1)>0$ gives the
contribution of the bound states to the total, bound and
scattering state probability. In a strict sense, correlated e-h
scattering states also contribute to the maximum of $g_{eh}(r)$,
but their influence is negligible for the temperatures considered,
where $r^b$ is of the order of several Bohr radii. For example,
for $T=50 K$ only excitons with the principle quantum number $n=1$
and $n=2$ are stable, which corresponds to a predominant
electron-hole separation in the range of about $4a_B$. This is
well reproduced by the procedure outlined above which yields $r^b
\simeq 3.7 a_B$, cf. Fig.~\ref{geehh} (upper left part). Hence the
fraction of the electron-hole bound states can be estimated from
the ratio
\begin{eqnarray}
\frac{N_{eh}^b}{N_{eh}^b+N_{eh}^c} =
\frac{\int_{0}^{r^b}r^2*[g_{eh}(r)-1]dr}{\int_{0}^{r^b}r^2*
g_{eh}(r)dr}\,,
 \label{alfar}
\end{eqnarray}
where the denominator is the normalization constant and has the
physical meaning of electrons to be in bound or scattering states.

Figure~\ref{cmpexp} shows the temperature dependence of the
fraction of electron-hole bound states calculated using
Eq.~(\ref{alfar}) for two dielectric constants and different
electron to hole mass ratios. Obviously, larger $\varepsilon$'s
result in lower electron-hole binding energies and therefore shift
the ionization temperature to smaller values. The same tendency is
observed if $m_h/m_e$ is lowered. For example, from
Fig.~\ref{cmpexp} we found that the ratio of the temperatures,
where the fraction of electron-hole bound states is $50\%$,  is
the same as the corresponding ratio of the reduced electron-hole
masses $\mu=m_e m_h/(m_e+m_h)$ (keeping $\varepsilon$ fixed). We
have a ratio of about 2 for the data belonging to $m_h=80$ and
$m_h=2.1$, and 1.5 for those belonging to $m_h=80$ and $m_h=4.2$.
This is exactly what we would expect from the ideal Saha equation
because the exciton binding energy is known to be proportional to
the reduced electron-hole mass.

\section{Conclusions}\label{dis}
To summarize, we have performed a largely unbiased direct path
integral quantum Monte Carlo simulation of a neutral non-ideal
two-component Coulomb system consisting of light electrons and
heavy holes. The careful analysis of typical many-particle
configurations, pair distribution and charge structure functions
reveals a rather comprehensive physical picture of how the system
behaves as a function of temperature in the low-density regime. We
obtained clear evidence for exciton formation at low temperatures
and a temperature-induced dissociation of these electron-hole
bound states. The ionization temperature strongly depends on the
dielectric constant and the mass ratio of electrons and holes,
where small values of $\varepsilon$ and large mass asymmetry
stabilize the bound states. As found out recently, these excitonic
bound states are the limiting factor for a possible Coulomb
crystallization of the heavy holes at low
densities~\cite{bfflf05}. At large temperatures the system
resembles an electron-hole plasma.

Electron-hole systems with such a strong mass anisotropy are
realized in intermediate valent $\rm Tm[Se_xTe_{1-x}]$
alloys~\cite{wachter}. In these materials f-d hybridization
provides us with a narrow dispersive f valence band and, as a
consequence, with a large effective hole mass of the order of
50-100 (bare) electron masses. $\rm TmSe_{0.45}Te_{0.55}$ is, at
ambient conditions, an indirect semiconductor having a gap of
$E_{\Delta}=130$~meV, where with $E_X^b\simeq 50-70$~meV below the
bottom of the d band an excitonic level has been observed.
Applying pressure the gap can be tuned (and even closed) and the
material is speculated to realize in the pressure region between 5
and 11~kbar an excitonic insulator, the search for which has been
run for a long time. A necessary precondition is the existence of
a large number of (up to $\sim 10^{20}$) excitons with
intermediate size (in order to avoid too strong overlap of the
excitonic bound states)~\cite{wachter}. Surprisingly, in $\rm
TmSe_{0.45}Te_{0.55}$ the excitonic phase then is predicted to be
stable at rather high temperatures (up to 200~K). Using the $\rm
Tm[Se_xTe_{1-x}]$ parameters within our (surely oversimplified)
two-component Coulomb model DPIMC simulation, we could at least
corroborate this belief. We found that (i) the fraction of
excitonic bound states amounts to 80-90~\%,  
(ii) the bound states will be stable up to
100-150~K, and (iii) the mass asymmetry between holes and electron
is crucial therefore.  As yet we are not in the position to
detect the condensed excitonic phase. Feasible, however, should be the
investigation of the density dependence of the various quantities.
Here we expect to see the formation of bi-exciton bound states and
finally the Mott transition at large particle densities $n_a$
(i.e. small $r_s$). Note that the excitonic-insulator semi-metal
(Mott) transition has been detected to occur in $\rm
TmSe_{0.45}Te_{0.55}$ at a pressure of about 13-14~kbar for $T\to
0$. Work along this line is in progress.
\newpage
\section*{Acknowledgements}
This work was supported by RF President Grant
No. MK-1769.2003.08, RAS program No. 17
and Deutsche Forschungsgemeinschaft through BO 1366/2 and
FE 398/6. Research was also partly sponsored by Award No. PZ-013-02
of the U.S. Civilian Research \& Development Foundation for the Independent
States of the Former Soviet Union (CRDF) and of Ministry of
Education of Russian Federation. The authors would like to thank
G. Schubert and G. Wellein for assistance by implementing the code
on diverse parallel compute clusters.\\


\begin{thebibliography}{10}
\bibitem{note1} Note that the distance between two electrons (two holes) is $d
2^{1/3}$ and the conventional Bruckner parameter of electrons
(holes) follows from our definition by multiplication with the
$2^{1/3}$.
\end{thebibliography}
\end{document}